\documentclass[aps,pre,reprint,superscriptaddress]{revtex4-2}

\usepackage{graphicx}
\usepackage{amsmath}
\usepackage{amsfonts}
\usepackage{bm}
\usepackage[separate-uncertainty]{siunitx}
\usepackage{xcolor}

\renewcommand{\vec}[1]{\bm{#1}}
\newcommand{\abs}[1]{\left\vert #1 \right\vert}

\newcommand{\order}[1]{\mathcal{O}\left\{#1\right\}}

\begin{document}

\title{Spectral holographic trapping:
Creating dynamic force landscapes with
polyphonic waves}

\author{Mia C. Morrell}
\affiliation{Department of Physics and Center for Soft Matter Research,
New York University, New York, NY 10003, USA}

\author{Julianne Lee}
\affiliation{Bronx High School of Science,
New York, NY 10468, USA}
\altaffiliation{current affiliation: Cornell University, Ithaca, NY 14853, USA}

\author{David G. Grier}
\affiliation{Department of Physics and Center for Soft Matter Research,
New York University, New York, NY 10003, USA}

\date{\today}

\begin{abstract}
Acoustic trapping uses forces exerted
by sound waves to transport small objects
along specified trajectories in three dimensions.
The structure of the acoustic force landscape
is governed by the amplitude and phase
profiles of the sound's pressure wave.
These profiles can be controlled through
deliberate spatial modulation of monochromatic waves,
by analogy to holographic optical trapping.
Alternatively, spatial and temporal control
can be achieved by interfering a small number
of sound waves at multiple frequencies to
create acoustic holograms based on spectral
content.
We demonstrate spectral holographic trapping
by projecting acoustic conveyor beams that
move millimeter-scale objects along prescribed
paths, and control the complexity of particle trajectories by tuning the strength of weak reflections.
Illustrative spectral superpositions of static
and dynamic force landscapes enable us to realize two variations on the theme of a wave-driven oscillator, a deceptively
simple dynamical system with surprisingly complex phenomenology.
\end{abstract}

\maketitle

Forces exerted by sound waves can levitate and transport small objects without physical contact, which is a boon for processing sensitive \cite{laurell2007chip} and hazardous materials \cite{foresti2013acoustophoretic}.
The interplay of sound waves with small scatterers also provides an archetypal model for investigating the physics of wave-matter composite systems \cite{burns1989optical,dholakia2010colloquium,lim2022mechanical,abdelaziz2021ultrasonic}.
Most implementations of acoustic trapping use
sound waves of a single fixed frequency and achieve dexterous
control by suitably structuring the waves' amplitude
and phase profiles with large arrays of acoustic ``pixels'' \cite{marzo2019holographic, Ochiai2014}.
Like holographic optical traps \cite{grier2003revolution},
this kind of acoustic trapping pattern
is reconfigured by actively changing the wavefront
structure at each pixel in the array.

Here, we draw attention to an alternative approach to
dynamic acoustic trapping whose ability to move
matter along prescribed paths
is encoded in the spectral content of a small number of acoustic sources.
We illustrate the potential utility of such
spectral traps by demonstrating
dynamic acoustic manipulation along
a single axis using just two acoustic pixels
emitting stationary sound fields.

\section{Acoustic Forces}

\begin{figure}
   \centering
   \includegraphics[width=\columnwidth]{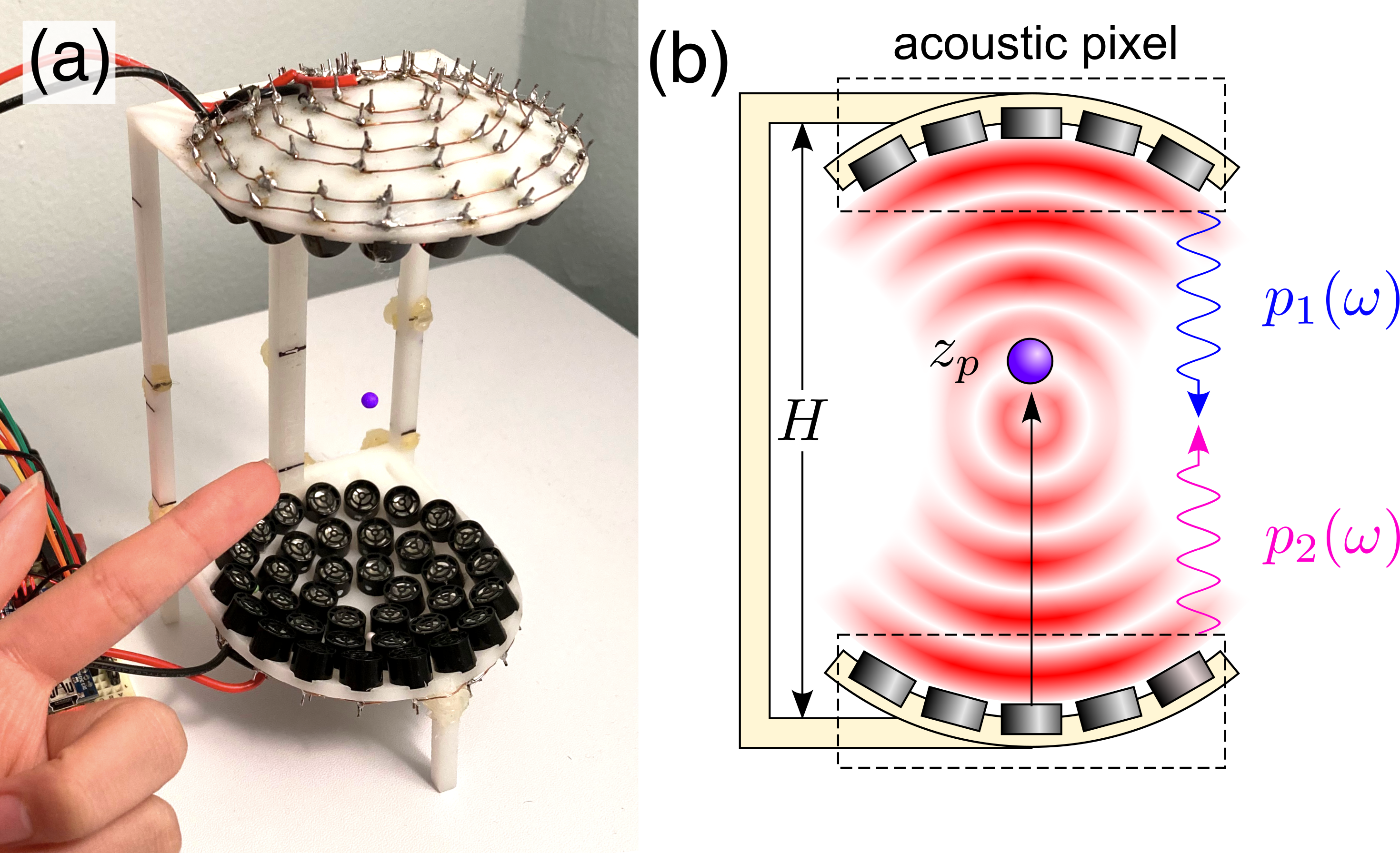}
    \caption{(a) Acoustic trap holding a single
    millimeter-scale particle in air at a carrier
    frequency of $f = \SI{40}{\kilo\hertz}$.
    (b) Schematic representation of spectral holographic
    trapping. Two acoustic pixels project counterpropagating pressure waves,
    $p_1(\omega)$ and $p_2(\omega)$,
    into a spherical cavity of height $H$.
    Dense rigid particles can be trapped at nodes
    in the combined pressure field.
    The traps' positions evolve in time based
    on the spectral content of the two projected
    waves.}
    \label{fig:intro}
\end{figure}

A steady sound wave at frequency $\omega$ propagating through an
incompressible medium can be characterized by
the real-valued amplitude profile,
$u(\vec{r})$,
and phase profile, $\phi(\vec{r})$,
of its pressure field:
\begin{equation}
\label{eq:pressure_field}
    p(\vec{r}, t)
    =
    u(\vec{r}) \,
    \exp\left(i \phi(\vec{r})\right) \,
    \exp(-i \omega t) .
\end{equation}
This structured wave can be decomposed
into plane waves whose wave number
satisfies the
standard dispersion relation,
$k = \omega / c_m$, where
$c_m$ is the speed of sound in the medium.
A small particle at position $\vec{r}$ in this field
experiences a time-averaged acoustic force \cite{Bruus2012Acoustofluidics2P,abdelaziz2020acoustokinetics}
\begin{subequations}
    \label{eq:acoustokinetic}
\begin{equation}
    \vec{F}(\vec{r})
    =
    - \nabla U_G(\vec{r}) + \order{(ka_p)^3},
\end{equation}
to leading order in the particle's
dimensionless size scale, $k a_p < 1$,
where
\begin{equation}
    \label{eq:gorkov}
    U_G(\vec{r})
    =
    \frac{1}{4} \, A \, u^2
    +
    \frac{1}{8k^2} \, B \, \nabla^2 u^2
\end{equation}
\end{subequations}
is the classic Gor'kov potential
\cite{bruus2012acoustofluidics,abdelaziz2020acoustokinetics}.
The force landscape described by
Eq.~\eqref{eq:acoustokinetic}
depends only on the pressure wave's
amplitude profile, and is is manifestly
conservative if $u(\vec{r})$ itself does
not depend on time.
The phase profile, $\phi(\vec{r})$,
governs nonconservative acoustic radiation forces \cite{abdelaziz2020acoustokinetics}
that may be neglected
for sufficiently small particles,
and vanish identically in standing waves.

For the specific case of a spherical particle
of radius $a_p$,
the coefficients $A$ and $B$ depend on the particle's density, $\rho_p$,
and sound speed, $c_p$, relative to those of the medium as
\cite{bruus2012acoustofluidics}
\begin{subequations}
    \label{eq:spherecoefficients}
\begin{align}
    \label{eq:alpha}
    A
    & =
    \frac{4 \pi}{3}a_p^3 \, \kappa_m
    \left(f_0 - \frac{3}{2} f_1\right) \quad \text{and} \\
    B
    \label{eq:beta}
    & =
    - 2\pi a_p^3 \, \kappa_m \, f_1,
\end{align}
where the monopole coupling coefficient,
\begin{equation}
    f_0 = 1 - \frac{\kappa_p}{\kappa_m} ,
\end{equation}
gauges the mismatch in
isentropic compressibility
between the particle, $\kappa_p = (\rho_p c_p^2)^{-1}$, and the medium, $\kappa_m = (\rho_m c_m^2)^{-1}$,
and the dipole coupling coefficient,
\begin{equation}
    f_2 = \frac{\rho_p - \rho_m}{\rho_p + \frac{1}{2} \rho_m} ,
\end{equation}
\end{subequations}
accounts for the mismatch in density.

\section{Dynamic Acoustic Trapping with Spectral Holograms}

A superposition of sound waves can be expressed in the form
of Eq.~\eqref{eq:pressure_field}, with $u(\vec{r})$ and
$\varphi(\vec{r})$ representing the amplitude and phase
of the associated interference pattern.
This approach has been used to create holographic acoustic traps
\cite{Melde2023compact,Melde2016hologram, Brown2019phase,marzo2019holographic,Fushimi2021acoustic,Marzo2015holographic},
typically by superposing waves
of the same frequency emanating from large arrays of sources.
Dynamic trapping patterns are created by
suitably modifying the amplitudes and phases of the individual sources over time \cite{Marzo2015holographic}.

Alternatively, time-varying acoustic force landscapes can be created by
superposing steady sound waves at different frequencies;
the resulting beats manifest
as slow time variations of the Gor'kov potential.
Equation~\eqref{eq:acoustokinetic} captures
this behavior by implicitly
averaging the acoustic force
over one period,
$T = 2 \pi/\omega$, of the carrier frequency, $\omega$.
The spectral content of a polyphonic superposition can
replace the spatiotemporal variations in a standard acoustic hologram
to create dynamic acoustic traps.
We refer to this frequency-based approach to wavefront shaping
as ``spectral holography''.

To illustrate the opportunities created by spectral holography,
we demonstrate programmable transport along one spatial dimension
using force landscapes created with just two acoustic pixels.
Our experimental system, illustrated in Fig.~\ref{fig:intro}, was introduced by Marzo, Barnes and Drinkwater \cite{marzo2017tinylev} and
consists of two banks of ultrasonic transducers operating
in air at a nominal frequency of \SI{40}{\kilo\hertz}.
Each bank acts as a single acoustic pixel, projecting
sound into a cylindrical section of a spherical cavity
of diameter $H = \SI{10}{\cm}$.
The counterpropagating waves interfere
within the cavity to create alternating nodes and antinodes of the
pressure field along the central axis.

\subsection{Diphonic Acoustic Conveyor}

If both sources operate at the same frequency, $\omega$,
their interference creates a standing wave with
axial nodes separated by half a wavelength.
Each node acts as a potential energy well for small
incompressible particles, such as the expanded polystyrene bead shown in Fig.~\ref{fig:intro}(a).

Detuning the two sources by
$\Delta \omega \ll \omega$ creates beats in the
axial pressure field,
\begin{align}
\label{eq:conveyorpressure}
    p(z, t)
    =
    2 p_0 \,
    \cos\left(kz - \frac{\Delta \omega}{2} t \right) \,
    \cos\left(\frac{\Delta \omega}{2 c_m} z - \omega t \right),
\end{align}
that manifest themselves as motion of the time-averaged axial force field,
\begin{equation}
\label{eq:conveyorforce}
    \vec{F}_a(z, t)
    =
    F_0(\omega) \,
    \sin(2kz - \Delta\omega \, t)\, \hat{z} ,
\end{equation}
after substitution into Eq.~\eqref{eq:acoustokinetic}.
The prefactor,
\begin{equation}
\label{eq:F0}
    F_0(\omega) = (A - 2B) \, k p_0^2,
\end{equation}
is positive for dense incompressible particles, which
therefore tend to be trapped at the nodes
of the pressure field.
The entire force landscape moves
along $\hat{z}$ at a steady speed,
\begin{equation}
\label{eq:conveyorvelocity}
    v_c = c_m \, \frac{\Delta \omega}{2 \omega},
\end{equation}
that is proportional to the detuning, $\Delta \omega$.
Setting aside complications
due to inertia and drag \cite{landau_fluids_book,settnes2012forces,baresch2016observation,morrell23},
trapped particles should travel
along with the landscape,
\begin{equation}
    z_p(t)
    \approx
    z_n(t)
\end{equation}
where $z_n(t) = z_n(0) + v_c t$ is the position of the
$n$-th pressure node at time $t$.
This type of traveling force landscape is known as a ``conveyor''
\cite{cizmar2005optical,cizmar2009tunable,ruffner2012optical,marzo2017tinylev,li2021acoustic,kandemir2021}
and is the simplest example
of a spectral hologram.

\begin{figure}
\centering
\includegraphics[width=0.9\columnwidth]{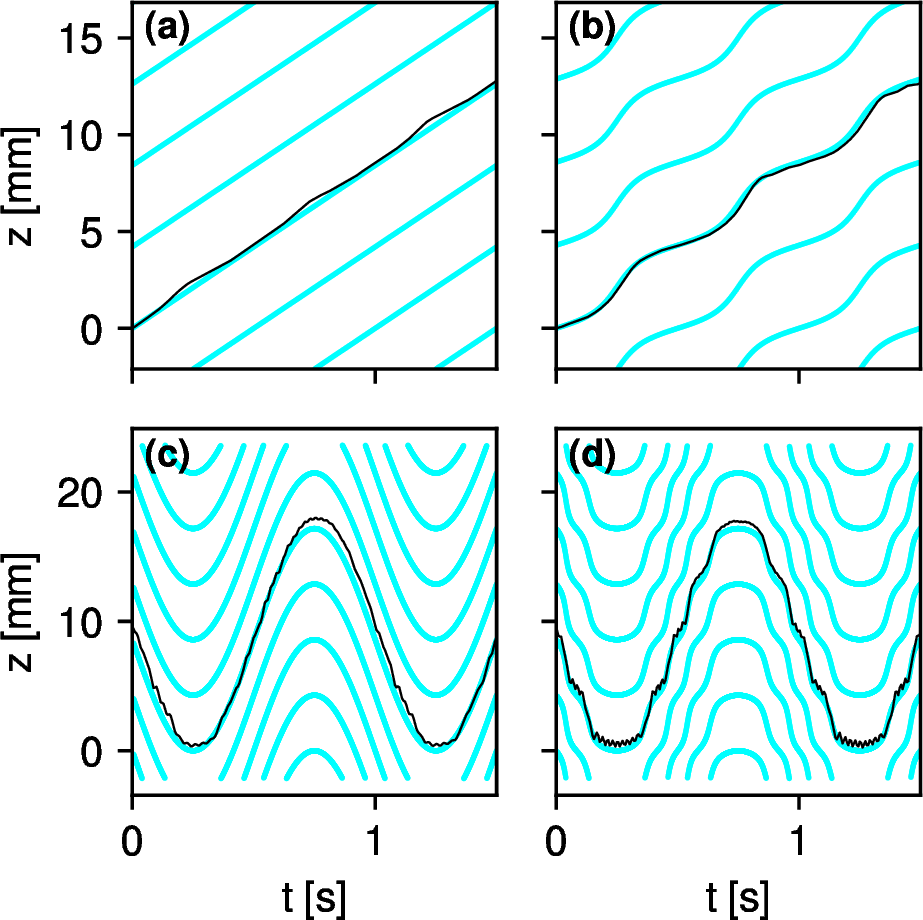}
\caption{Measured trajectories (black curves)
for a \SI{2}{\mm}-diameter EPS bead
(a) and (b) in an acoustic conveyor,
and (c) and (d) in an acoustic scanner.
These are
compared with predictions of Eq.~\eqref{eq:trap_traj} for
$\epsilon_0 = \num{0.38}$
 (cyan curves).
(b) and (d) Tuning the carrier frequency to a cavity resonance
at $f = \omega/(2 \pi) = \SI{40.0(1)}{\kilo\hertz}$
creates a standing wave that modulates the trajectory.
(a) and (c) Detuning to $f = \SI{40.7}{\kilo\hertz}$
suppresses the standing
wave.
Conveyor:
$\Delta f = \Delta \omega/(2 \pi) = \SI{2}{\hertz}$.
Scanner:
$\Delta \phi = \SI{720}{\pi}$.}
\label{fig:groovy}
\end{figure}

The data in Fig.~\ref{fig:groovy}(a) demonstrate an acoustic conveyor
transporting a millimeter-scale bead composed of type II expanded
polystyrene (EPS) foam with a measured
\cite{morrell23} mass density of
$\rho_p = \SI{30.5(2)}{\kg\per\cubic\meter}$.
The particle's trajectory is recorded at
\SI{170}{frames\per\second} with a
monochrome video camera
(FLIR, Blackfly S USB3)
whose \SI{5}{\ms} exposure time is fast
enough to avoid motion blurring given the system
magnification of \SI{61}{\um\per pixel}.
Each frame in a video sequence is thresholded with Otsu's method, and
the particle's position is computed as the center of mass of
the resulting simply-connected cluster of
foreground pixels.
The image of a typical particle
yields a \num{1000}-pixel cluster
whose axial centroid, $z_p(t)$, can be located with
an estimated accuracy \cite{morrell23} of
$\Delta z_p = \SI{0.17}{pixel} = \SI{10}{\um}$,
which suffices for our application.

\subsection{Polyphonic Acoustic Scanner}

More sophisticated modes of transport
can be achieved with more sophisticated
superpositions of tones.
Such a generalized conveyor, which we call a ``scanner",
can be implemented as the superposition
of waves from two sources, as depicted
in Fig.~\ref{fig:intro}(b),
with time-varying relative phase,
\begin{subequations}
\label{eq:generalized_spectral_pressure}
\begin{align}
    p_1(t)
    & =
    p_0 \, e^{i \omega t} \\
    p_2(t)
    & =
    p_0 \, e^{i\varphi(t)} \, e^{-i \omega t}.
\end{align}
\end{subequations}
Such a superposition creates
a time-averaged force landscape,
\begin{equation}
    \vec{F}_a(z, t)
    =
    F_0(\omega) \,
    \sin(2kz - \varphi(t))\, \hat{z},
\end{equation}
whose traps travel along $\hat{z}$ as
\begin{equation}
\label{eq:generalizedvelocity}
    z_n(t)
    =
    z_n(0) - \frac{1}{2k} \, \varphi(t).
\end{equation}
The resulting motion is slow in the sense
that relevant variations in the relative phase
satisfy $\abs{\dot \varphi} \ll \omega$,
where the dot refers to a derivative
with respect to time.
Any
faster variations are suppressed in theory
by the implicit time average in
Eq.~\eqref{eq:acoustokinetic}
and physically by
viscous drag and the particle's inertia.

Active control of the relative phase, $\varphi(t)$,
has been used in the context of
holographic optical trapping to project
optical conveyors \cite{cizmar2005optical} and
optical tractor beams \cite{ruffner2012optical},
and more recently has been used
to demonstrate acoustic conveyors \cite{marzo2017tinylev}.
Rather than actively sweeping the phase, however,
we instead can decompose $\varphi(t)$
into its spectral components and use those
to create a scanner that operates in
steady state without active intervention.
For example, a sinusoidal scanner described by
$\varphi(t) = \Delta \phi \, \sin(\Omega t)$
can be implemented through the
Jacobi-Anger identity,
\begin{equation}
    p_2(t)
    =
    p_0 \, \sum_{n = -\infty}^{\infty} J_n(\Delta\phi)\, e^{i(n\Omega-\omega)t},
\end{equation}
which specifies the
frequencies needed
to implement the scanner and their relative
amplitudes.
A working example can be
projected with just
the first few orders, $n \in [-4, 4]$.
The resulting spectral hologram then
transports trapped objects back and
forth continuously and smoothly without active intervention.
The data in Fig.~\ref{fig:groovy}(c) show
such a scanner in action.

\subsection{Spectral Superposition of Static and Dynamic Landscapes}

Acoustic pixels are actively driven transducers.
As a consequence, they not only project
sound waves, but also act as
absorbing boundary conditions
for incident waves \cite{mokry2003sound}.
This feature has not been emphasized
in previous acoustic-trapping
studies \cite{marzo2017tinylev}.
Active cancellation of reflections
enables the counterpoised acoustic pixels
in an instrument
such as the example in Fig.~\ref{fig:intro}
to create standing-wave
acoustic traps even when the carrier
frequency, $\omega$, is not
tuned to a cavity resonance.

In practice, acoustic
pixels reflect a small proportion, $\epsilon_0$, of incident sound
waves.
Reflections contribute
to the force landscape by
forming standing waves within
the cavity whose amplitude can be controlled
through the choice of $\omega$.
The associated force landscape therefore
has both time-varying and stationary components,
\begin{subequations}
\label{eq:spectralsuperposition}
\begin{equation}
\label{eq:reflection_force}
\vec{F}(\vec{r}, t)
=
\vec{F}_a(\vec{r}, t) +
\vec{F}_s(\vec{r}),
\end{equation}
where the standing-wave contribution is
approximately
\begin{equation}
\label{eq:static_force_landscape}
\vec{F}_s(\vec{r})
\approx
2 \,
\epsilon(\omega) \, F_0(\omega) \, \sin(2kz) \, \hat{z} .
\end{equation}
The factor of \num{2}
in Eq.~\eqref{eq:static_force_landscape} accounts for
the independent contributions from
each of the pixels.
The depth of the stationary landscape's
modulation,
\begin{equation}
\label{eq:epsilon}
    \epsilon(\omega)
    =
    \epsilon_0 \,
    \cos\bigg(\frac{2H}{c_m} \omega \bigg),
\end{equation}
\end{subequations}
is proportional to the acoustic pixels' reflection coefficient, $\epsilon_0$,
and can be tuned by adjusting the carrier
frequency.
For the cavity depicted in Fig.~\ref{fig:intro}, we find that
particle trajectories are consistent
with $\epsilon_0 = \num{0.38(2)}$.
The overall scale of the stationary force
landscape is set by $F_0(\omega)$,
which is given by Eq.~\eqref{eq:F0}.

\begin{figure}
    \includegraphics[width=0.9\columnwidth]{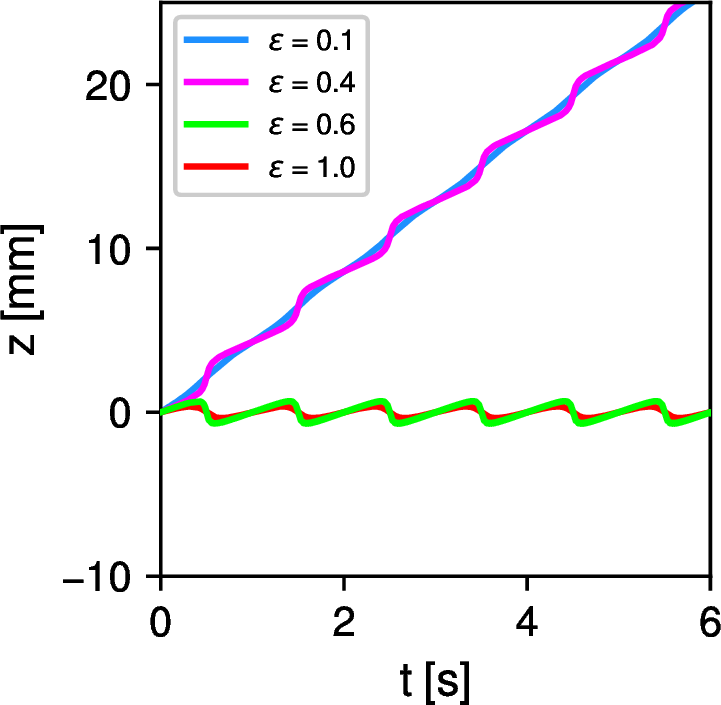}
    \caption{Simulated trajectories of the particle
    from Fig.~\ref{fig:groovy} in
    a  spectrally superposed acoustic conveyor, Eq.~\eqref{eq:spectralsuperposition}, as a function
    of the cavity reflection coefficient,
    $\epsilon(\omega)$.
    Conveyor detuning: $\Delta f =
    \Delta \omega / (2 \pi) = \SI{1}{\hertz}$.}
    \label{fig:sims}
\end{figure}

If the reflection coefficient is large enough,
$\epsilon_0 > 0.5$, the central frequency
can be tuned so that $2\epsilon(\omega) > 1$.
In that case, the standing wave exerts enough force to
trap the particle, and the
dynamic landscape acts as a time-dependent perturbation.
Representative trajectories for this mode of motion
are plotted in Fig.~\ref{fig:sims} as a function of $\epsilon(\omega)$.

In the opposite limit of weak reflections, $\epsilon(\omega) < 0.5$,
the particle is transported by the moving conveyor
across the stationary landscape.
The nodes then trace out trajectories,
\begin{equation}
\label{eq:trap_traj}
    z_n(t \vert \epsilon(\omega))
    =
    z_n(0) -
    \frac{1}{k} \, \arctan\left(
    \frac{\epsilon(\omega)-1}{\epsilon(\omega)+1}
    \tan\left(\frac{\varphi(t)}{2}
    \right)
    \right),
\end{equation}
that reduce to Eq.~\eqref{eq:generalizedvelocity} when $\epsilon(\omega) = 0$.
This mode of motion also is plotted in
Fig.~\ref{fig:sims}
and is consistent with the perturbed
trajectories
observed experimentally in
Fig.~\ref{fig:groovy}(b)
and Fig.~\ref{fig:groovy}(d).

The particles' trajectories increasingly deviate
from the traps' trajectories as
$\epsilon(\omega)$ approaches
$1/2$ and the trajectories become
increasingly sinuous.
We quantify these deviations with
the kinematic variance,
\begin{equation}
\label{eq:kinematic_variance}
    \sigma^2(\omega)
    =
    \frac{1}{T}
    \int_0^T
    [z_p(t) - z_n(t)]^2 \, dt ,
\end{equation}
which is plotted as a function
of the carrier frequency, $\omega$,
in Fig.~\ref{fig:conv}(a).
Measurements are compared with
the prediction obtained
by setting $z_p(t) = z_n(t \vert \epsilon(\omega))$, which is plotted
as a solid curve.
The kinematic model is consistent with the measurement when the carrier frequency
is tuned away from the cavity resonance
so that the particle travels smoothly
at constant speed.
Tuning to the cavity resonance at
$f = \SI{40.2}{\kilo\hertz}$
maximizes the particle's acceleration
and increases deviations between
the trajectories of the particle and
the trap.
These discrepancies can be resolved by
accounting for inertial corrections
to the viscous drag acting
on the particle.

\section{Accounting for Inertia and Drag}

\begin{figure*}
    \includegraphics[width=0.9\textwidth]{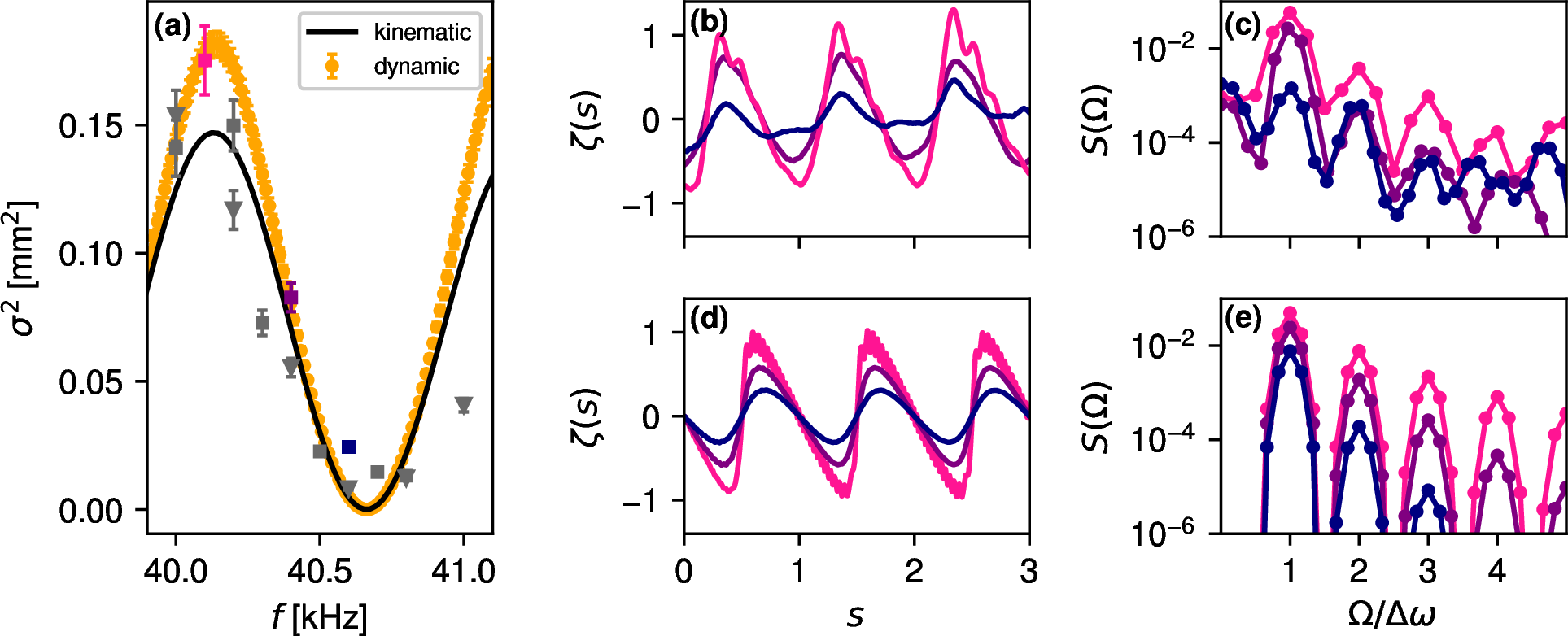}
    \caption{
    (a) Variance of excursions from the mean conveyor trajectory as a function of carrier frequency. Measurements
    are plotted as discrete points. Solid curves denote analytic predictions of Eq.~\eqref{eq:trap_traj} (black)
    and numerical solutions to Eq.~\eqref{eq:wavedrivenoscillator}
    (orange).
    (b) Typical measured trajectories
    colored to match corresponding
    points in (a).
    (c) Power spectra of the trajectories in (b) showing the growth of harmonics in the particle's trajectory as
    the depth of modulation increases.
    (d) and (e) numerical solutions
    of Eq.~\eqref{eq:wavedrivenoscillator}
    for the same set of conditions.
    Conveyor detuning frequency:
    $\Delta f = \Delta \omega/2 \pi = \SI{2}{\hertz}$.
     }
    \label{fig:conv}
\end{figure*}

A trapped particle hews to the
trajectory of its acoustic
trap if the motion is
slow enough to neglect
the inertia of the fluid medium
\cite{morrell23}.
More generally, the equation of motion
for a particle of mass $m_p$,
\begin{equation}
\label{eq:generalequationofmotion}
    m_p \ddot{z}_p
    =
    F_a(z_p, t) +
    F_s(z_p) +
    F_d(\dot{z}_p, \ddot{z}_p),
\end{equation}
reflects contributions from
the active force landscape,
the stationary force landscape
and viscous drag, respectively.
A sphere of radius $a_p$
moving through a fluid of viscosity
$\eta_m$ and mass density
$\rho_m$ experiences a drag force
that is described by the
Basset-Boussinesq-Oseen equation
\cite{maxey1983equation,landau_fluids_book,
lovalenti1993hydrodynamic,baresch2016observation},
\begin{subequations}
    \label{eq:drag}
\begin{widetext}
\begin{equation}
\label{eq:dragforce}
    F_d(\dot{z}_p, \ddot{z}_p)
    =
    6 \pi \eta_m a_p
    \left[
    \dot{z}_p +
    \tau \ddot{z}_p
    +
    \sqrt{\frac{9\tau}{\pi}}
    \int_{-\infty}^t
    \frac{\ddot{z}_p(t')}{\sqrt{t - t'}}
    \, dt'
    \right],
\end{equation}
\end{widetext}
which accounts for the inertia of the
displaced fluid on time scales set
by the viscous relaxation time,
\begin{equation}
    \label{eq:tau}
    \tau
    =
    \frac{\rho_m}{9 \eta_m} \, a_p^2.
\end{equation}
\end{subequations}
The history-dependent contribution to the
drag complicates an analytic formulation
of the transport properties for
a general spectral hologram.
To illustrate the challenge,
we consider the comparatively
simple case of
a particle moving under
the influence of an acoustic conveyor.
Competition between the
active and stationary force
landscapes causes
the particle to oscillate
at the beat frequency, $\Delta \omega$, about the moving trap's
position.
We therefore define the dimensionless
displacement in the co-moving
frame,
\begin{equation}
\label{eq:zeta}
    \zeta(t)
    =
    2 k z_p(t) -
    \begin{cases}
	\Delta\omega \, t,
        & \epsilon(\omega) < 0.5 \\
        0,
        & \epsilon(\omega) > 0.5
    \end{cases} .
\end{equation}
Applying Eq.~\eqref{eq:generalequationofmotion} and Eq.~\eqref{eq:drag}
then yields the deceptively simple
dimensionless equation of
motion,
\begin{equation}
    \label{eq:wavedrivenoscillator}
    \zeta''
    + b \, \zeta'
    + \zeta
    =
    \tilde{\epsilon}
    \sin\left(
    \zeta -
    \frac{\Delta \omega}{\omega_0} \, s \right),
\end{equation}
where primes denote derivatives
with respect to the dimensionless time,
$s = \omega_0 t$.
Equation~\eqref{eq:wavedrivenoscillator}
describes a wave-driven oscillator
\cite{abdelaziz2021dynamics} whose exceptionally
rich phenomenology only recently has been
brought to light.
Wave-driven oscillators differ from
more familiar nonlinear dynamical systems,
such as the Duffing oscillator
\cite{andrade2014nonlinear,fushimi2018nonlinear}, because its spatial nonlinearity
is irreducibly coupled to the time dependence of the driving.

The effective driving strength in Eq.~\eqref{eq:wavedrivenoscillator},
\begin{equation}
    \tilde{\epsilon}(\omega)
    =
    \begin{cases}
	2 \epsilon(\omega),
        & \epsilon(\omega) < 0.5 \\
        \frac{1}{2\epsilon(\omega)},
        & \epsilon(\omega) > 0.5
    \end{cases} ,
\end{equation}
can be varied over the
range $\tilde{\epsilon}(\omega)
\in [0, 1]$ by adjusting the carrier
frequency relative to the cavity
resonance.
Similarly, the natural frequency,
\begin{equation}
    \label{eq:omega0}
    \omega_0(\omega)
    =
    \sqrt{\frac{2kF_0}{m}} \times
    \begin{cases}
	1 ,
        & \epsilon(\omega) < 0.5 \\
        \sqrt{2 \epsilon(\omega)} , & \epsilon(\omega) > 0.5
    \end{cases} ,
\end{equation}
and the drag coefficient,
\begin{equation}
\label{eq:b}
    b(\omega)
    =
    \frac{6 \pi \eta_m a_p}{m \, \omega_0} \times
    \begin{cases}
        1 ,
        & \epsilon(\omega) < 0.5 \\
        (2 \epsilon(\omega))^{-1},
        & \epsilon(\omega) > 0.5
    \end{cases},
\end{equation}
both depend on cavity tuning when
$\epsilon(\omega) > 0.5$.

Equations~\eqref{eq:omega0} and \eqref{eq:b}
incorporate the inertial
corrections from Eq.~\eqref{eq:drag}
by introducing the dynamical mass
\cite{landau_fluids_book,settnes2012forces,morrell23},
\begin{subequations}
\label{eq:inertialcorrection}
\begin{equation}
\label{eq:dynamicmass}
    m(\Delta \omega)
    =
    m_p \left\{
    1 + \frac{1}{2} \frac{\rho_m}{\rho_p}
    \left[ 1
    + \frac{9}{2} \frac{\delta(\Delta \omega)}{a_p} \right]
    \right\},
\end{equation}
under the simplifying assumption
 the particle oscillates harmonically at the driving frequency,
$\Delta \omega$.
The sphere's effective mass is increased
in this approximation
by the mass of the fluid in a
Prandtl-Schlichting boundary layer
of thickness
\cite{landau_fluids_book}
\begin{equation}
\label{eq:delta}
    \delta(\Delta \omega)
    =
    \sqrt{\frac{2 \eta_m}{\rho_m}
    \frac{1}{\Delta \omega}} .
\end{equation}
\end{subequations}
This correction has been demonstrated
to quantitatively model the damped
oscillations of a particle levitated
in a static acoustic trap
\cite{morrell23}.
For particles moving in an
acoustic conveyor, the dynamic
model more accurately accounts
for the magnitude of measured
fluctuations, as can be seen
in Fig.~\ref{fig:conv}(a).

Measured acoustic-conveyor trajectories in
Fig.~\ref{fig:conv}(b)
and their power spectra in Fig.~\ref{fig:conv}(c)
are reproduced reasonably well by the
numerical solutions
of Eq.~\eqref{eq:wavedrivenoscillator}
that are plotted in
Fig.~\ref{fig:conv}(d) and Fig.~\ref{fig:conv}(e).
These examples illustrate the effect of
tuning the carrier frequency on the amplitude
and harmonic content of the particle's dynamic response.
Values of $F_0$ and $m_0$ used for the numerical
solutions are obtained from measured trajectories
using the analytical approach
described in Ref.~\cite{morrell23}.
The power spectra are computed as
\begin{equation}
    S(\Omega)
    =
    \abs{
    \int_0^1 \zeta(s) \, W(s) \,
    e^{-i \Omega s} \,
    ds
    }^2,
\end{equation}
using the Blackman-Harris window
function, $W(s)$.
The wave-driven oscillator
responds most strongly
at the driving frequency,
$\Omega = \Delta \omega$.
Increasingly much power
is directed into
harmonics of that driving
frequency as the depth
of modulation increases.
Agreement between the measured and computed power spectra
illustrates the
utility of the Basset-Boussinesq-Oseen equation for interpreting the behavior of wave-driven
oscillators created with sound.
At the same time, the presence of strong harmonics
suggests even better agreement could be attained
by seeking self-consistent solutions to the equation
of motion, including the BBO correction
described in Eq.~\eqref{eq:drag}.

More generally, the wave-driven oscillator
has been shown \cite{abdelaziz2021dynamics} to respond
at both harmonics and subharmonics
of the driving frequency and to undergo
transitions between subharmonic states
depending on the strength of the
driving, $\tilde{\epsilon}(\omega)$,
the strength of the damping,
$b(\omega)$
and the relationship between the
driving frequency, $\Delta \omega$,
and the oscillator's
natural frequency, $\omega_0$.
Transitions between subharmonic
states feature both period-doubling
routes to chaos and Fibonacci cascades
\cite{abdelaziz2021dynamics}.
No complete description of the
wave-driven oscillator is yet available,
even in the weak-driving regime, $\tilde{\epsilon} < 1$.
Previous experimental and numerical
studies \cite{abdelaziz2021dynamics},
furthermore,
have neglected the inertial corrections described by Eq.~\eqref{eq:drag}
that are likely to have influenced
their results.
Future studies of the wave-driven oscillator
and related dynamical systems would benefit
both from the streamlined experimental
implementation afforded by spectral holography
and also from the analytical approach
discussed here.

\section{Discussion}

Spectral holographic trapping
uses interference among waves at
multiple frequencies to create time-averaged
force landscapes that evolve dynamically on
the inertial time scales of trapped objects.
Spatiotemporal control afforded by the frequency
content of the projected waves reduces the complexity
of acoustic manipulation systems by replacing the
many spatial degrees of freedom
required for conventional monotonic
holographic projection.
Spectral holography therefore allows complex force
landscapes to be generated with small numbers of acoustic pixels.
We have demonstrated two archetypal examples,
a unidirectional conveyor created with two frequencies and a bidirectional scanner created with nine.
We also have shown that tuning the carrier frequency
to cavity resonances
can usefully implement a superposition of dynamic
and static force fields with no additional complexity.
In the case of an acoustic conveyor, this superposition
implements a wave-driven oscillator whose exceedingly rich
dynamical properties emerge from an interplay between
the acoustic force field, the particle's inertia
and viscous drag in the supporting medium.
This study also highlights the importance of
accounting for the fluid's inertia when planning and
interpreting the motions of particles in
acoustic force landscapes.

The combination of rich spectral control and
analytic dynamical modeling expands the prospects
for dexterous acoustic manipulation of macroscopic
materials.
The present study has focused on the dynamics of
individual particles in spectral holograms
created within cavities.
Additional opportunities can be imagined for
free-space manipulation with traveling waves
and for self-organization guided by wave-mediated
interactions in many-body systems immersed in
spectral holograms.

\begin{acknowledgments}
This work was supported by the National Science Foundation under Award Number DMR-2104837.
\end{acknowledgments}

%

\end{document}